\documentclass[12pt]{article} 
\usepackage{amssymb}
\usepackage{amstext}
\usepackage{latexsym}

 \newtheorem{theorem}{Theorem}
 \newtheorem{lemma}[theorem]{Lemma}
 	
\newcommand{\Z}{{\mathbb{Z}}}
\newcommand{\F}{{\mathbb{F}}}
\newcommand{\R}{{\mathbb{R}}}

\newcommand{\CC}{{\mathcal{C}}}

\newcommand{\E}{{\mathsf{E}}}

\newcommand{\RR}{{\mathcal{R}}}

\newcommand{\Eb}{\mathbf E}

\newcommand{\Nb}{\mathbf N}

\newcommand{\Sb}{\mathbf S}

\newcommand{\ub}{\mathbf u}
\newcommand{\Ub}{\mathbf U}

\newcommand{\Vb}{\mathbf V}

\newcommand{\xb}{\mathbf x}
\newcommand{\Xb}{\mathbf X}

\newcommand{\Yb}{\mathbf Y}

\newcommand{\Zb}{\mathbf Z}

\newcommand{\lambdab}{\mbox{\boldmath$\lambda$}}

\newcommand{\eps}{\varepsilon}

\newcommand{\ie}{{\em i.e., }}
\newcommand{\eg}{{\em e.g., }}

\newcommand{\qed}{\hspace*{1cm}\hspace*{\fill}\openbox}
\newcommand{\half}{\frac{1}{2}}

\newcommand{\mod}{~\mathrm{mod}~}  
\newcommand{\SNR}{\mathrm{SNR}}  

\newcommand{\openbox}{\leavevmode
  \hbox to.77778em{%
  \hfil\vrule
  \vbox to.675em{\hrule width.6em\vfil\hrule}%
  \vrule\hfil}}
\newcommand{\proofname}{Proof}

\begin{document}
\renewcommand{\textfraction}{0}

\title{On the role of MMSE estimation \\ in approaching the
information-theoretic limits \\ of linear Gaussian channels: \\ Shannon
meets Wiener}
 \author{\normalsize
G. David Forney, Jr.\footnote{I am grateful to J. M. Cioffi, U.
Erez, R. Fischer and R. Zamir for many helpful comments.}
 \\[-5pt]
\small MIT \\[-5pt]Ê
\small Cambridge, MA 02139 USA \\[-5pt] \small
forneyd@comcast.net }
\date{}
\maketitle
\thispagestyle{empty}
\begin{abstract}
We discuss why MMSE estimation arises in
lattice-based schemes for approaching the capacity of linear Gaussian
channels, and comment on its properties.
\end{abstract}Ê
\normalsize

\section{Introduction}

Recently, Erez and Zamir \cite{EZ01, ZSE02} have cracked
the long-standing problem of achieving the capacity of additive white
Gaussian noise (AWGN) channels using lattice codes and lattice
decoding.  Their method uses Voronoi codes (nested lattice codes),
dither, and an MMSE estimation factor $\alpha$ that had previously been
introduced in more complex multiterminal scenarios, such as Costa's
``dirty-paper channel" \cite{C83}.  However, they give no fundamental
explanation for why an MMSE estimator, which is seemingly an artifact
from the world of analog communications, plays such a key role in the
digital communications problem of achieving channel capacity.

The principal purpose of this paper is to provide such an explanation, in
the lattice-based context of a  mod-$\Lambda$ AWGN channel model.  We
discuss various properties of MMSE-based schemes in this application, some
of which are unexpected. 

MMSE estimators also appear as part of capacity-achieving solutions for
more general linear Gaussian channel scenarios;
 \eg in MMSE-DFE structures (including precoding) for ISI channels
\cite{EF92, CDEFI95}, and
generalized MMSE-DFE structures for vector and multi-user channels
\cite{CF97,YC01}. 
Some of the
explanation for the ``canonicality" of MMSE-DFE structures in the these
more general scenarios is no doubt information-theoretic
\cite{SL96, GV98}.  The observations of this paper
complement these results by showing why lattice-type codes combine so
well with MMSE equalization structures, as shown previously in
\cite{GV00,ZSE02}.

\section{Lattice-based coding for the AWGN channel} 

Consider the real discrete-time AWGN channel $Y = X + N$, 
where $\E[X^2] \le S_x$ and $N$ is independent\footnote{
Note that without the independence of $N$, the ``additive" property is
vacuous, since for any real-input, real-output channel we may define
$N = Y - X$, and then express $Y$ as $Y = X + N$.  We exploit this
idea later.} zero-mean
Gaussian noise with variance $S_n$.  The
capacity is $C = \half \log_2 (1 + \SNR)$ bits per dimension (b/d),
where $\SNR = S_x/S_n$.  Following Erez and Zamir \cite{EZ01,
ZSE02}, we will show how lattice-based transmission systems can approach
the capacity of this channel at all SNRs. 

\subsection{Lattices and spheres}

Geometrically, an $N$-dimensional lattice $\Lambda$ is a regular infinite
array of points in $\R^N$.  Algebraically, $\Lambda$ is a
discrete subgroup of $\R^N$ which spans $\R^N$. A Voronoi region
$\RR_V(\Lambda)$ of $\Lambda$ represents
the quotient group $\R^N/\Lambda$ by a set of minimum-energy coset
representatives for the cosets of $\Lambda$ in $\R^N$.  For any $\xb \in
\R^N$, ``$\xb \mod
\Lambda$" denotes the unique element of $\RR_V(\Lambda)$ in the coset
$\Lambda + \xb$.  Geometrically, $\R^N$ is the disjoint union of the
translated Voronoi regions $\{\RR_V(\Lambda) + \lambdab, \lambdab \in
\Lambda\}$.  The volume $V(\Lambda)$ of $\RR_V(\Lambda)$ is therefore the
volume of $\R^N$ associated with each point of $\Lambda$.

As $N \to \infty$, the Voronoi regions of some $N$-dimensional lattices
can become more or less spherical, in various senses.  
As $N \to \infty$, an $N$-sphere (ball)
of squared radius $N\rho^2$ has normalized volume (per two dimensions)
$$
V_\otimes(N\rho^2)^{2/N} \stackrel{N \to \infty}{\longrightarrow}
2 \pi e \rho^2.
$$
The average energy per dimension of a uniform probability distribution
over such an $N$-sphere goes to $P_\otimes(N\rho^2) = \rho^2$.  The
probability that an iid Gaussian random $N$-tuple with zero mean and
symbol variance $S_n$ falls outside the $N$-sphere becomes arbitrarily
small for any $S_n < \rho^2$.

It is known that there exist high-dimensional lattices whose Voronoi
regions are quasi-spherical in the following second moment sense.  The 
\emph{normalized second moment} of a compact region $\RR \subset \R^N$ of
volume $V(\RR)$ is defined as
$$
G(\RR) =  \frac{P(\RR)}{V(\RR)^{2/N}},
$$
where $P(\RR)$ is the average energy per dimension of a uniform
probability distribution over $\RR$.  The normalized second moment of
$\RR$ exceeds that of an $N$-sphere. The normalized second moment of an
$N$-sphere decreases monotonically with $N$ and approaches
$\frac{1}{2\pi e}$ as $N \to \infty$.  Poltyrev (reported in
Feder-Zamir \cite{ZF96}) showed that there exist lattices $\Lambda$
such that $\log 2 \pi e G(\Lambda)$ is arbitrarily small, where
$G(\Lambda)$ denotes the normalized second moment of $\RR_V(\Lambda)$. 
Such lattices are said to be ``good for quantization," or ``good for
shaping."

Poltyrev \cite{P94} also showed that there exist high-dimensional
lattices whose Voronoi regions are quasi-spherical in the sense that the
probability that an iid Gaussian noise $N$-tuple with symbol variance
$S_n$ falls outside the Voronoi region $\RR_V(\Lambda)$ is arbitrarily
small as long as
$$
S_n < \frac{V(\Lambda)^{2/N}}{2 \pi e}.
$$
Such lattices are said to be ``good for AWGN channel coding," or
``sphere-bound-achieving" \cite{FTC00}.

\subsection{Mod-lattice transmission and capacity}

We now show that the mod-$\Lambda$ transmission
system shown in Figure 1 can approach the channel capacity
$C = \half \log_2 (1 + S_x/S_n)$ b/d arbitrarily closely,
provided that $G(\Lambda) \approx 1/(2\pi e)$ and $f(\Yb)$ is a MMSE
estimator of $\Xb$.  

\setlength{\unitlength}{5pt}
\begin{center}
\begin{picture}(67,10)
\put(0,2){\vector(1,0){15}}
\put(2,3){$\Xb \in \RR_V(\Lambda)$}
\put(6,0){$S_x$}
\put(17,2){\circle{4}}
\put(16,1.5){$+$}
\put(17,8){\vector(0,-1){4}}
\put(18,5){$S_n$}
\put(16,9){$\Nb$}
\put(19,2){\vector(1,0){11}}
\put(24,3){$\Yb$}
\put(30,0){\framebox(4,4){$f$}}
\put(34,2){\vector(1,0){10}}
\put(37,3){$f(\Yb)$}
\put(44,0){\framebox(8,4){mod $\Lambda$}}
\put(52,2){\vector(1,0){15}}
\put(54,3){$\Yb' \in \RR_V(\Lambda)$}
\end{picture}
\end{center}
\begin{center}
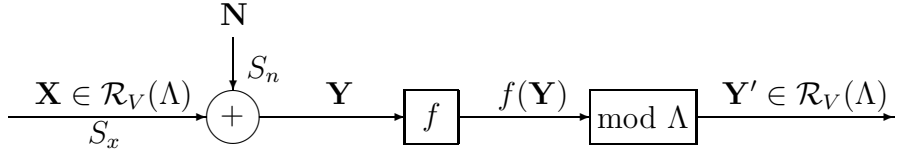
Figure 1.  Mod-$\Lambda$ transmission system over an AWGN channel.
\end{center}

This system is based on an $N$-dimensional lattice $\Lambda$ whose Voronoi
region
 $\RR_V(\Lambda)$ has volume $V(\Lambda)$, average energy per dimension
$P(\Lambda) = S_x$ under a uniform probability distribution over
$\RR_V(\Lambda)$, and thus normalized second moment $G(\Lambda) = 
P(\Lambda)/V(\Lambda)^{2/N}$.

The $N$-dimensional input vector $\Xb$ is restricted to the Voronoi
region $\RR_V(\Lambda)$.  The output vector $\Yb$ is mapped by some
function $f$ to another vector $f(\Yb) \in \R^N$, which is then
mapped modulo $\Lambda$ to $\Yb' = f(\Yb) \mod \Lambda$, also in the
Voronoi region
$\RR_V(\Lambda)$.


Our main result is that capacity can be approached in the system of
Figure 1 if and only if the lattice $\Lambda$ is ``good for shaping" and
the function $f(\Yb)$ is an MMSE estimator.  (The sufficiency of these
conditions was shown in \cite{EZ01, ZSE02}.)

As a first step, we derive a lower bound:

\begin{theorem}[Mod-$\Lambda$ channel capacity]
The capacity
 $C(\Lambda, f)$ of the mod-$\Lambda$ transmission system of
Figure 1 is lowerbounded by
$$
C(\Lambda, f) \ge C - \half \log_2 2\pi
e G(\Lambda) -\half \log_2 \frac{S_{e,f}}{S_e} \quad \mbox{b/d},
$$ 
where $C = \half \log_2 (1 + \SNR)$ b/d is the capacity of the underlying
AWGN channel, $G(\Lambda)$ is the normalized second moment of
$\RR_V(\Lambda)$, and $S_{e,f}$ and $S_e$ are the average energies per
dimension of $\Eb_f = f(\Yb) - \Xb$ and of $\Eb = \hat{\Xb}(\Yb) - \Xb$,
respectively, where $\hat{\Xb}(\Yb)$ is the linear MMSE estimator of $\Xb$
given $\Yb$.
\end{theorem} 

The key to the proof of this theorem is the introduction of a dither
variable $\Ub$ that is known to both transmitter and receiver, and whose 
probability distribution is uniform over the Voronoi region
$\RR_V(\Lambda)$, as in \cite{EZ01, ZSE02}.  Given a data
vector
$\Vb \in \RR_V(\Lambda)$, the channel input is taken as
$$
\Xb = \Vb + \Ub \mod \Lambda.
$$
This makes $\Xb$ a uniform random variable over $\RR_V(\Lambda)$,
\emph{statistically independent of} $\Vb$.  
This property follows from the following lemma:\footnote{
We call this the crypto lemma because if we take $X$ as plaintext,
$N$ as a cryptographic key, and $Y = X + N$ as the encrypted
message, then the encrypted message is independent of the plaintext
provided that the key is uniform, so no information can be obtained about
the plaintext from the encrypted message without the key.  On the
other hand, given the key, the plaintext may be easily recovered
from the encrypted message via $X = Y - N$.  This is the principle of the
one-time pad, which, as Shannon showed, is essentially the only way to
achieve perfect secrecy in a cryptographic system.}

\begin{lemma}[Crypto lemma]  Let $G$ be a compact abelian group\footnote{
The group $G$ is required to be compact so that its Haar
(translation-invariant) measure $\mu(G)$ is finite and thus
normalizable to a uniform probability distribution over $G$.  However,
$G$ need not be abelian.}
 with group
operation $+$, and let $Y = X + N$, where $X$ and $N$ are random
variables over $G$ and $N$ is independent of $X$ and uniform over $G$. 
Then $Y$ is independent of $X$ and uniform over $G$. 
\end{lemma} 
\emph{Proof}.  Since $y-x$ runs through $G$ as $y$ runs
through $G$ and $p_N(n)$ is constant over $n \in G$, the distribution
$p_{Y|X}(y|x) = p_N(y - x)$ is constant over $y \in G$ for any $x \in G$.
\qed

One effect of the dither $\Ub$ is thus to ensure that the channel
input $\Xb = \Vb + \Ub \mod \Lambda$ is uniform over $\RR_V(\Lambda)$ and
thus has average energy per dimension $P(\Lambda) = S_x$.  A second and
more important effect is to make $\Xb$ and thus also $\Yb = \Xb + \Nb$
independent of $\Vb$.

The dither may be subtracted out at the output of the channel, mod
$\Lambda$, to give
$$
\Zb = f(\Yb) - \Ub \mod \Lambda.
$$ 
  The
end-to-end channel is illustrated in Figure 2.

\setlength{\unitlength}{5pt}
\begin{center}
\begin{picture}(86,10)
\put(0,2){\vector(1,0){12}}
\put(0,3){$\Vb \in \RR_V(\Lambda)$}
\put(14,2){\circle{4}}
\put(13,1.5){$+$}
\put(14,8){\vector(0,-1){4}}
\put(13,9){$\Ub$}
\put(16,2){\vector(1,0){2}}
\put(18,0){\framebox(8,4){mod $\Lambda$}}
\put(26,2){\vector(1,0){4}}
\put(27,3){$\Xb$}
\put(32,2){\circle{4}}
\put(31,1.5){$+$}
\put(32,8){\vector(0,-1){4}}
\put(31,9){$\Nb$}
\put(34,2){\vector(1,0){4}}
\put(35,3){$\Yb$}
\put(38,0){\framebox(4,4){$f$}}
\put(42,2){\vector(1,0){7}}
\put(43,3){$f(\Yb)$}
\put(51,2){\circle{4}}
\put(50,1.5){$+$}
\put(51,8){\vector(0,-1){4}}
\put(49,9){$-\Ub$}
\put(53,2){\vector(1,0){2}}
\put(55,0){\framebox(8,4){mod $\Lambda$}}
\put(63,2){\vector(1,0){23}}
\put(64,3){$\Zb = f(\Yb)- \Ub \mod \Lambda$}
\end{picture}
\end{center}
\begin{center}
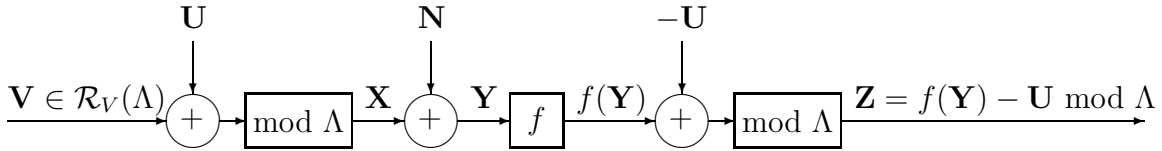
Figure 2.  Creation of a mod-$\Lambda$ channel $\Zb =
f(\Yb)- \Ub \mod \Lambda$ using dither.
\end{center}

Now let us regard $f(\Yb)$ as an estimator of $\Xb$, and define the
estimation error as $\Eb_f = f(\Yb) - \Xb$.  Since $\Yb$ and $\Xb$ are
independent of $\Vb$, so is $\Eb_f$.  Then
$$
\Zb = \Xb + \Eb_f - \Ub = \Vb + \Eb_f \mod \Lambda.
$$ 
In short, we have created a mod-$\Lambda$ additive noise channel $\Zb =
\Vb + \Eb_f \mod \Lambda$, where $\Eb_f$ is independent of $\Vb$.  This
equivalent channel is illustrated in Figure 3.

\setlength{\unitlength}{5pt}
\begin{center}
\begin{picture}(56,10)
\put(0,2){\vector(1,0){12}}
\put(0,3){$\Vb \in \RR_V(\Lambda)$}
\put(14,2){\circle{4}}
\put(13,1.5){$+$}
\put(14,8){\vector(0,-1){4}}
\put(13,9){$\Eb_f$}
\put(16,2){\vector(1,0){10}}
\put(17,3){$\Vb + \Eb_f$}
\put(26,0){\framebox(8,4){mod $\Lambda$}}
\put(34,2){\vector(1,0){22}}
\put(35,3){$\Zb = \Vb + \Eb_f \mod \Lambda$}
\end{picture}
\end{center}
\begin{center}
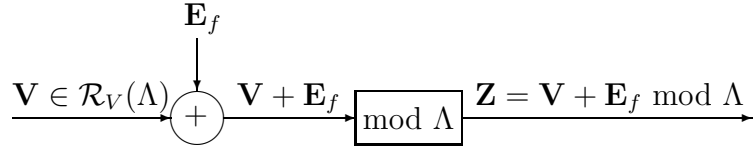
Figure 3.  Equivalent mod-$\Lambda$ additive noise channel $\Zb =
\Vb + \Eb_f \mod \Lambda$.
\end{center}

As is well known, the capacity of an additive-noise channel $\Zb =
\Vb + \Eb_f \mod \Lambda$ is achieved when the input
distribution is uniform over $\RR_V(\Lambda)$, in which case the output
distribution is uniform as well, by the crypto lemma. The capacity is
equal to 
$$
C(\Lambda, f) = \frac{1}{N}(h(\Zb) - h(\Zb \mid \Vb)) =
\frac{1}{N}(\log_2 V(\Lambda) - h(\Eb_f')) \quad \mbox{b/d},
$$
where $h(\Zb) = \log_2 V(\Lambda)$ is the differential entropy of a
uniform distribution over a region of volume $V(\Lambda)$, and
$h(\Eb_f')$ is the differential entropy of the $\Lambda$-aliased additive
noise $\Eb_f' = \Eb_f \mod \Lambda$.  Now since $\Eb_f'$ is the result of
applying the many-to-one mod-$\Lambda$ map to $\Eb_f$, we have 
$$
h(\Eb_f') \le h(\Eb_f).
$$
Moreover, if $\Eb_f$ has average energy per dimension $S_{e,f}$, then we
have 
$$
h(\Eb_f) \le \frac{N}{2} \log_2 2 \pi e S_{e,f},
$$
the differential entropy of an iid zero-mean Gaussian distribution with
the same average energy.  Combining these results,  using 
$V(\Lambda)^{2/N} = P(\Lambda)/G(\Lambda)$ and $P(\Lambda) = S_x$, we have
$$
C(\Lambda, f) \ge \frac{1}{N}\log_2 V(\Lambda) - \half \log_2 2 \pi e
S_{e,f} = \half \log_2 \frac{S_x}{S_{e,f}} - \half \log_2 2 \pi e
G(\Lambda)
\quad
\mbox{b/d}.
$$

The linear MMSE estimator $\hat{\Xb}(\Yb)$ of $\Xb$ is $\hat{\Xb}(\Yb) =
\alpha \Yb$, where
$$
\alpha = \frac{S_x}{S_x + S_n} = \frac{\SNR}{1 + \SNR}.
$$ 
By the orthogonality principle of MMSE estimation theory, the linear MMSE
estimation error $\Eb = \Xb - \alpha\Yb = (1 - \alpha)\Xb -
\alpha \Nb$ is then uncorrelated with $\Yb$.\footnote{
These relations are illustrated by the ``Pythagorean" right triangle
shown in Figure 4 below, which follows from interpreting covariances as
inner products of vectors in a two-dimensional Hilbert space.  Since
$\E[XN] = 0$, the two vectors corresponding to $X$ and $N$ are
orthogonal.  Their squared lengths are given by $\E[X^2] = S_x$ and
$\E[N^2] = S_n$.  The hypotenuse corresponds to the sum $Y = X + N$, and
has squared length
$S_y = S_x + S_n$.  Since $\E[XY] = S_x$, the projection of
$Y$ onto $X$ is $\hat{Y}(X) = (\E[XY]/\E[X^2])X = X$, and the projection
of $X$ onto $Y$ is  $\hat{X}(Y) = (\E[XY]/\E[Y^2])Y = \alpha Y$.
Then $E = X - \hat{X}(Y) = X - \alpha Y$ is orthogonal to $Y$.
The inner right  triangle in Figure 4 with sides
$(\hat{X}(Y), E, X)$ is  similar, so since $S_x
= \alpha S_y$ the squared lengths of its sides are $(S_{\hat{x}} =
\alpha S_x, S_e = \alpha S_n, S_x = \alpha S_y)$, respectively.  


\setlength{\unitlength}{5pt}
\begin{center}
\begin{picture}(30,17)(0,1)
\put(0,2){\line(1,0){25}}
\put(25,2){\line(0,1){18.75}}
\put(0,2){\line(4,3){25}}
\put(25,2){\line(-3,4){9}}
\put(9,0){$X: S_x$}
\put(26,11){$N: S_n$}
\put(9,14){$Y: S_y$}
\put(19,6){$E$}
\put(7,6){$\hat{X}(Y) = \alpha Y$}
\end{picture}
\end{center}
\begin{center}
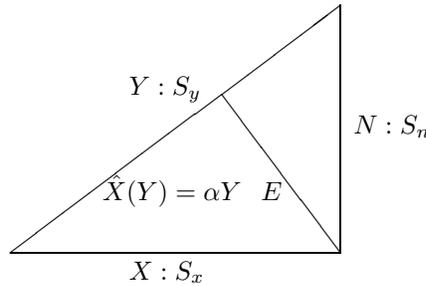
Figure 4.   ``Pythagorean" right triangle with sides $(X, N, Y)$, \\
with similar inner right triangle with sides  $(\hat{X}(Y) = \alpha Y, E
=  X - \hat{X}(Y), X)$.  
\end{center}
}  
The average energy of the estimation error per dimension becomes
$$
S_e = (1 - \alpha)^2 S_x + \alpha^2 S_n = \frac{S_x S_n}{S_x + S_n} =
\alpha S_n
$$
(see footnote). Finally, since $S_x/S_e = S_y/S_n = 1 + \SNR$, we have
$$
C(\Lambda, f) \ge \half \log_2 \frac{S_x}{S_{e}} + \half \log_2
\frac{S_{e}}{S_{e,f}} - \half \log_2 2 \pi e G(\Lambda) = C - \half \log_2
2
\pi e G(\Lambda) - \half \log_2
\frac{S_{e,f}}{S_{e}}  \quad \mbox{b/d}.
$$
This completes the proof of Theorem 1.
\qed

\textbf{Remark 1} (dither is unnecessary).  Evidently a
channel $\Zb = \Vb + \ub +\Eb_f \mod \Lambda$ with a fixed dither vector
$\ub \in \RR_V(\Lambda)$ has the same capacity $C(\Lambda, f)$.  Therefore
introducing the random dither variable $\Ub$  is just a tactic to
prove Theorem 1; dither is not actually needed to achieve
$C(\Lambda, f)$.  However, dither is key to decoupling the Shannon and
the Wiener problems.\footnote{This is analogous to the tactic used by
Elias \cite{E55} to prove that binary linear block codes can achieve the
capacity of a binary input-symmetric channel, namely the introduction of
a random translate $\CC + \Ub$ of a binary linear block code $\CC$ of
length $N$, where $\Ub$ is a random uniform binary $N$-tuple
in $(\F_2)^N$.}
\qed

\textbf{Remark 2} (MMSE estimation and bias).  Notice that the
signal-to-noise ratio of the channel $\Zb = \Vb + \Eb \mod
\Lambda$ is
$S_x/S_e = S_y/S_n = 1 + \SNR$,
and moreover this channel has no bias.  Thus the MMSE factor $\alpha$ and
random dither increase the effective signal-to-noise ratio
from $\SNR$ to $1 + \SNR$ without introducing bias.  This is evidently a
different way of approaching capacity than that given in \cite{CDEFI95},
where the apparent $\SNR_{\mathrm{MMSE-DFE}}$ was discounted to
$\SNR_{\mathrm{MMSE-DFE,U}} = \SNR_{\mathrm{MMSE-DFE}} - 1$ to account
for bias.  
\qed

\textbf{Remark 3} (``dirty-paper" capacity).  This approach easily
extends to give a constructive proof of Costa's result \cite{C83} that
channel interference known to the transmitter does not reduce capacity; 
see, \eg \cite{ZSE02, BCW03}.  Let the channel model be $\Yb = \Xb +
\Nb +
\Sb$, where $\Sb$ is an arbitrary interference vector known to the
transmitter.  Then let the channel input be $\Xb = \Vb + \Ub - \alpha\Sb
\mod
\Lambda$.  The channel input is still uniform and independent of $\Vb$,
by the crypto lemma, while the effect of the interference $\Sb$ is
entirely cancelled in
$\Zb = \alpha \Yb - \Ub = \Vb + \Eb_f \mod \Lambda$.  Thus the receiver
needs to know nothing about the interference, the equivalent channel
model is the same, and $C(\Lambda, f)$ is unaffected.
\qed

Theorem 1 implies that the capacity $C = \half
\log_2(1+\SNR)$ can be approached arbitrarily closely by the mod-$\Lambda$
channel of Figure 1 if $\log 2 \pi e G(\Lambda) \to 0$ and
$f(\Yb)$ is the linear MMSE estimator $\hat{\Xb}(\Yb) = \alpha \Yb$,
which is the main result of Erez and Zamir \cite{EZ01}.


We now show that the conditions $\log_2 2 \pi e G(\Lambda)
\to 0$ and  $S_{e,f} = S_e$ are not only sufficient but also necessary to
reach capacity.  Briefly, the arguments are as follows:

  1.  The differential entropy per dimension of $\Xb$ and $\Zb$, namely
$$
\frac{1}{N}h(\Xb) = \frac{1}{N}h(\Zb) = \half \log_2 V(\Lambda)^{2/N} =
\half \log_2 2\pi e S_x - \half \log_2 2\pi e G(\Lambda)
$$
goes to $\half \log_2 2\pi e S_x$ if and only if $\log_2 2 \pi e
G(\Lambda) \to 0$.
This condition is necessary because the capacity of an
AWGN channel with input power constraint $S_x$ can be approached
arbitrarily closely only if $h(\Xb)/N$ approaches $\half \log_2 2\pi e
S_x$.

\textbf{Remark 4} (Gaussian approximation principle).  The differential
entropy of any random vector $\Xb$ with average energy per dimension
$S_x$ is less than or equal to $\half\log_2 2\pi e S_x$, with equality if
and only if $\Xb$ is iid Gaussian.  Therefore if $\Xb_n$ is a sequence of
random vectors of dimension $N(n) \to \infty$ and average energy per
dimension $S_x$ such that $h(\Xb_n)/N(n) \to \half\log_2 2\pi e S_x$, we
say that the sequence $\Xb_n$ is \emph{Gaussian in the limit}.  Restating
the above argument, if $\Xb_n$ is uniform over $\RR_V(\Lambda_n)$, then 
$\Xb_n$ is Gaussian in the limit if and only if 
$\log 2 \pi e G(\Lambda_n) \to 0$.\footnote{
Zamir and Feder \cite{ZF96} show that if $\Xb_n$ is uniform over an
$N(n)$-dimensional region $\RR_n$ of average energy $S_x$ and $G(\RR_n)
\to 1/(2\pi e)$, then the normalized divergence $\frac{1}{N(n)}D(\Xb_n ||
\Nb_n) \to 0$, where $\Nb_n$ is an iid Gaussian random vector with zero
mean and variance $S_x$.  They go on to show that this implies that any
finite-dimensional projection of $\Xb_n$ converges in distribution to an
iid Gaussian vector.}
\qed

  2.  The channel output $\Yb = \Xb + \Nb$ is then also Gaussian in the
limit, so the linear MMSE estimator $\hat{\Xb}(\Yb) = \alpha \Yb$ becomes
a true MMSE estimator in the limit.  The MMSE estimation error $\Eb =
-(1-\alpha)\Xb + \alpha\Nb$ becomes Gaussian in the limit with
symbol variance $S_e = \alpha S_n$, and becomes independent of $\Yb$.  In
order that $C(\Lambda, f) \to C$, it is then necessary that $S_{e,f} =
S_e$, which by definition implies that $f(\Yb)$ is an MMSE
estimator.\footnote{ Since $\Eb_f = \Eb + (f(\Yb) - \hat{\Xb}(\Yb))$ and
$\Yb$ and $\Eb$ are independent, $S_{e,f} = S_e + \frac{1}{N}\E[||f(\Yb) -
\hat{\Xb}(\Yb)||^2]$.  Thus $f(\Yb)$ is an MMSE estimator if and only if 
$\E[||f(\Yb) - \hat{\Xb}(\Yb)||^2] = 0$.}
\pagebreak

In summary, these two conditions are necessary as well as sufficient:
 
\begin{theorem}[Necessary conditions to approach $C$]
The capacity of the mod-$\Lambda$ 
\linebreak channel of Figure 1
approaches $C$ if and only if $\log 2\pi e G(\Lambda) \to 0$ and $f(\Yb)$
is an MMSE estimator of $\Xb$ given $\Yb$.
\end{theorem}

\textbf{Remark 5} (MMSE estimation and lattice decoding).  One
interpretation of the Erez-Zamir result is
that the scaling introduced by the MMSE estimator is somehow essential for
lattice decoding of a fine-grained coding lattice $\Lambda_c$.  Theorem 3
shows however that in the mod-$\Lambda$ channel an MMSE estimator is
necessary to achieve capacity, quite apart from any particular coding and
decoding scheme.
\qed

\textbf{Remark 6} (aliasing becomes negligible).  Under these conditions,
Theorem 1 says that $C(\Lambda, f) \ge C$.  Since $C(\Lambda, f)$ cannot
exceed $C$, this implies that all inequalities in the proof of Theorem 1
must tend to equality, and in particular that 
$$
\frac{h(\Eb')}{N}  \to \frac{h(\Eb)}{N}  \to \half \log_2 2 \pi e
S_e,
$$
where $\Eb' = \Eb \mod \Lambda$ is the $\Lambda$-aliased version of
the estimation error $\Eb$.  So not only must $\Eb$ become Gaussian in
the limit, \ie $h(\Eb)/N \to \half \log_2 2 \pi e S_e$,
but also $\Eb'$ must tend to $\Eb$, which means that the effect of the
mod-$\Lambda$ aliasing must become negligible.  This is as
expected, since $\Eb$ is Gaussian in the limit with symbol variance
$S_e$ and $\RR_V(\Lambda)$ is quasi-spherical with average energy
per dimension $S_x > S_e$.
\qed


\subsection{Voronoi codes}

A \emph{Voronoi code} $\CC((\Lambda_c+\ub)/\Lambda) = (\Lambda_c + \ub)
\cap \RR_V(\Lambda)$ is the set of points in a translate $\Lambda_c + \ub$
of an
$N$-dimensional ``coding lattice"
$\Lambda_c$ that lie in the Voronoi region $\RR_V(\Lambda)$ of a
``shaping" sublattice
$\Lambda \subset \Lambda_c$.  (Such codes were called ``Voronoi codes" in
\cite{CS83}, ``Voronoi constellations" in \cite{F89}, and ``nested
lattice codes" in \cite{EZ01, ZSE02, BCW03}.  Here we
will use the original term.)

A Voronoi code has $|\Lambda_c/\Lambda| = V(\Lambda)/V(\Lambda_c)$ code
points, and thus rate
$$
R(\Lambda_c/\Lambda) = \frac{1}{N} \log_2 \frac{V(\Lambda)}{V(\Lambda_c)}
= \half \left(\log_2 \frac{V(\Lambda)^{2/N}}{2 \pi e} - \log_2
\frac{V(\Lambda_c)^{2/N}}{2 \pi e}\right) \quad \mbox{b/d}.
$$

Erez and Zamir \cite{EZ01, ZSE02} have shown rigorously (not
employing the Gaussian approximation principle) that there exists a
random ensemble
$\CC((\Lambda_c + \Ub)/\Lambda)$ of dithered Voronoi codes that can
approach the capacity
$C(\Lambda)$ of the mod-$\Lambda$ transmission system of Figure 1
arbitrarily closely, if
$f(\Yb) = \hat{\Xb}(\Yb) = \alpha \Yb$.  The decoder may be the usual
minimum-Euclidean-distance decoder, even though the effective noise $\Eb =
-(1-\alpha)\Xb + \alpha\Nb$ is not Gaussian.
 
If $C(\Lambda) \approx C$ and $P(\Lambda) =
S_x$, this implies that $2 \pi e G(\Lambda) \approx 1$; \ie
$\Lambda$ is ``good for shaping."  Furthermore, since the effective noise
has variance $S_e$, if the error probability is arbitrarily small and
$R(\Lambda_c/\Lambda)
\approx C =
\half \log_2 S_x/S_e$, then 
$$
\log_2 S_e \approx \log_2 \frac{V(\Lambda_c)^{2/N}}{2 \pi e};
$$
\ie $\Lambda_c$ is ``good for AWGN channel coding," or
``sphere-bound-achieving."

The ensemble $\CC((\Lambda_c + \Ub)/\Lambda)$ is an ensemble
of fixed-dither Voronoi codes $\CC((\Lambda_c + \ub)/\Lambda)$.  The
average probability of decoding error $\Pr_\Ub(E) = \E_\Ub[\Pr_\ub(E)]$
is arbitrarily small over this ensemble, using a decoder that
is appropriate for random dither (\ie minimum-distance decoding). This
implies not only that there exists at least one fixed-dither code
$\CC((\Lambda_c + \ub)/\Lambda)$ such that
$\Pr_\ub(E) \le \Pr_\Ub(E)$, using the same decoder, but also that at
least a fraction $1- \eps$ of the fixed-dither codes have
$\Pr_\ub(E) \le \frac{1}{\eps}\Pr_\Ub(E)$;  \ie almost all fixed-dither
codes have low $\Pr_\ub(E)$.

This result is somewhat counterintuitive, since for fixed dither $\ub$,
$\Xb$ is not independent of $\Vb$;  indeed, there is a one-to-one
correspondence given by $\Xb = \Vb + \ub \mod \Lambda$.  Therefore, the
error
$$
\Eb = -(1-\alpha)\Xb + \alpha \Nb = -(1-\alpha)(\Vb + \ub \mod \Lambda) +
\alpha \Nb
$$
is not independent of $\Vb$;  \ie there is bias in the equivalent channel
output $\Zb = \Vb + \Eb \mod \Lambda$.  Even so, we see that
capacity can be achieved by a suboptimum decoder which ignores bias.

Since almost all fixed-dither codes achieve capacity, we may as well use
the code $\CC((\Lambda_c+\ub)/\Lambda)$ that has minimum average energy
$S_{\mathrm{min}} \le P(\Lambda) = S_x$ per dimension.  But if
$S_{\mathrm{min}} < S_x$, then we could achieve a rate greater than the
capacity of an AWGN channel with signal-to-noise ratio
$S_{\mathrm{min}}/S_n < S_x/S_n$.  We conclude that the average energy
per dimension of $\CC((\Lambda_c+\ub)/\Lambda)$ cannot be materially less
than $S_x = P(\Lambda)$ for any $\ub$, and thus  must be approximately
$S_x$ for almost all values of the dither $\ub$, in order for the average
over $\Ub$ to be $S_x$.
In summary:

\begin{theorem}[Average energy of Voronoi codes]
If $\CC((\Lambda_c+\ub)/\Lambda)$ is a capacity-achieving Voronoi code,
then $\Lambda_c$ is good for AWGN channel coding, $\Lambda$ is good for
shaping, the decoder may ignore bias, and the average energy per dimension
of $\CC((\Lambda_c+\ub)/\Lambda)$ is $\approx P(\Lambda)$. 
\end{theorem}

\textbf{Remark 7} (Average energy of Voronoi
codes). Theorem 4 shows that the hope of \cite{FTC00} that one
could find particular Voronoi codes with average energy 
$S_x - S_e$ was misguided.
For Voronoi codes, the original ``continuous approximation" of
\cite{FW89} holds, not the ``improved continuous approximation" of
\cite{FTC00}.

\textbf{Remark 8} (observations on output scaling).
It is surprising that a decoder for Voronoi codes which first scales the
received signal by $\alpha$ and then does lattice decoding should perform
better than one that just does lattice decoding.  Optimum (ML) decoding
on this channel is minimum-distance (MD) decoding, and ordinary lattice
decoding is equivalent to minimum-distance decoding except on the
boundary of the support region.

Scaling by $\alpha$ seems excessive.  
Scaling the output by
$\alpha$ reduces the received variance to $S_{\hat{x}} = \alpha^2 S_y =
\alpha S_x$, less than the input variance.  This means that the scaled
output $\alpha Y$ is almost surely going to lie in a spherical shell of
average energy per dimension
$\approx \alpha S_x$, whereas the code vectors in the Voronoi code
$\CC(\Lambda_c/\Lambda)$ almost all lie on a spherical shell of average
energy $\approx S_x$.  Yet the subsequent lattice decoding to
$\CC((\Lambda_c+\ub)/\Lambda)$ works, even though it seems that the
decoder should decode to $\alpha \CC((\Lambda_c+\ub)/\Lambda)$.

These questions about scaling may be resolved if as $N \to \infty$ it
suffices to decode Voronoi codes based on angles, ignoring magnitudes. 
Then whether the decoder uses $Y, \alpha Y $ or $\sqrt{\alpha} Y$ as
input, the optimum minimum-angle decoder would be the same.  Indeed,
Urbanke and Rimoldi \cite{UR98}, following Linder \emph{et al.}
\cite{LSZ93}, have shown that as $N \to \infty$ a suboptimum decoder for
spherical lattice codes that does minimum-angle decoding to the subset of
codewords in a spherical shell of average energy $\approx S_x$ suffices
to approach capacity. 

Of course, lattice decoding does depend on scale, so it seems that
scaling the lattice decoder is just a trick to analyze the optimal
minimum-angle decoder performance, as well as to show that lattice
decoding of Voronoi codes suffices to reach capacity.

Finally, note that with a fixed code and scaling by $\alpha$, as $N \to
\infty$ the output $\alpha\Yb$ almost surely lies in a sphere of average
energy $\approx \alpha S_x < S_x$, inside 
$\RR_V(\Lambda)$, so the mod-$\Lambda$ operation in the receiver has
negligible effect and may be omitted.
\qed

%

\textbf{Remark 9} (Shannon codes, spherical lattice codes, and Voronoi
codes).  In Shannon's random code ensemble for the AWGN channel, the code
point $\Xb$ asymptotically lies almost surely in a spherical shell of
average energy per dimension $\approx S_x$, the received vector
$\Yb$ lies almost surely in a spherical shell of average energy per
dimension
$\approx S_y$, and the noise vector $\Nb$ lies almost surely in a
spherical shell of average energy per dimension $\approx S_n$. 
Thus we obtain a geometrical picture in which a ``output sphere" of
average energy
$\approx S_y$ is partitioned into $\approx (S_y/S_n)^{N/2}$
probabilistically disjoint ``noise spheres" of squared radius $\approx
S_n$.  Curiously, the centers of the noise spheres are at average energy
$\approx S_x$, even though practically all of the volumes of the noise
spheres are at average energy $\approx S_y$.

Urbanke and Rimoldi \cite{UR98} have shown that spherical lattice codes
(the set of all points in a lattice $\Lambda_c$ that lie within a sphere
of average energy $S_x$) can achieve the channel capacity $C = \half
\log_2 S_y/S_n$ b/d with minimum-distance decoding.  Since again $\Yb$ and
$\Nb$ must lie almost surely in spheres of average energy $S_y$ and
$S_n$, respectively, we again have a picture in which the output sphere
must be partitioned into $\approx (S_y/S_n)^{N/2}$ effectively disjoint
noise spheres whose centers are the points in the spherical lattice code,
which have average energy $\approx S_x$.

Voronoi codes evidently work differently.  The Voronoi
region $\RR_V(\Lambda)$ has average energy $S_x$, and so does any good
Voronoi code $\CC((\Lambda_c+\ub)/\Lambda)$.  Moreover, $\RR_V(\Lambda)$
is the disjoint union (mod $\Lambda$) of $V(\Lambda)/V(\Lambda_c) \approx
(S_x/S_e)^{N/2}$ small Voronoi regions, whose centers are the points in 
$\CC((\Lambda_c+\ub)/\Lambda)$.  So \emph{the centers have the same
average energy as the bounding region}, in contrast to the spherical
case.  

By the sphere bound \cite{FTC00, P94} $\log_2
V(\Lambda_c)^{2/N}/(2 \pi e) \ge \log_2 S_c$, where $S_c$ is the channel
noise variance, so the capacity of the mod-$\Lambda$ channel is limited to
$\half \log_2 S_x/S_c$.  If the channel noise has variance $S_c = S_n$,
then the capacity is limited to $\overline{C} = \half \log_2 S_x/S_n =
\half
\log_2 \SNR$, which is the best that de Buda and others \cite{deB89, L97}
were able to achieve with Voronoi codes prior to \cite{EZ01}.  However,
the MMSE estimator reduces the effective channel noise variance to
$S_c = S_e  = \alpha S_n$, which allows the capacity to approach $C =
\half \log_2 S_x/S_e = \half \log_2 (1 + \SNR)$.  So in the mod-$\Lambda$
setting the MMSE estimator is the crucial element that precisely
compensates for the Voronoi code capacity loss from $C$ to the ``lattice
capacity" $\overline{C}$.  

Finally, consider a ``backward-channel" view of the Shannon ensemble. 
The jointly Gaussian pair $(X, Y)$ is equally
well modeled by the forward-channel model $Y = X + N$ or the
backward-channel model $X = \alpha Y + E$.  From the latter perspective,
the transmitted codeword $\Xb$ lies almost surely in a spherical shell of
average energy $\approx S_e$ about the scaled received word
$\alpha Y$, which lies almost surely in a spherical shell of
average energy $\approx \alpha^2 S_y = \alpha S_x$.    Thus we
obtain a geometrical picture in which an ``input sphere" of average
energy $\approx S_x$ is partitioned into $\approx (S_x/S_e)^{N/2}$
probabilistically disjoint ``decision spheres" of squared radius $\approx
S_e$.  The centers of the decision spheres are codewords of average energy
$\approx S_x$.

Capacity-achieving Voronoi codes thus appear to be designed according to
the backward-channel view of the Shannon ensemble, whereas
capacity-achieving spherical lattice codes appear to be designed
according to the forward-channel view.
\pagebreak

\footnotesize{

}
\end{document}